\documentclass[preprintnumbers,floatfix,aps,prd,preprint]{revtex4}

\usepackage{amsmath}
\usepackage{amsfonts}
\usepackage{amssymb}
\usepackage{bm}
\usepackage{enumerate}
\usepackage[dvips]{color,graphicx}
\usepackage{epsfig}
\usepackage[figuresright]{rotating}

\begin{document}
\preprint{JLAB-THY-11-7}

\title{Observation of quark-hadron duality in
	$\gamma^* p$ helicity cross sections}

\author{S.~P.~Malace$^1$, W.~Melnitchouk$^2$ and A.~Psaker$^3$}
\affiliation{
$^1$\mbox{Department of Physics, Duke University,
	Durham, North Carolina 27708, USA}	\\
$^2$Jefferson Lab, Newport News, Virginia 23606, USA \\
$^3$American University of Nigeria, Yola, Nigeria}

\begin{abstract}
Combining data on unpolarized and polarized inclusive proton structure
functions, we perform the first detailed study of quark-hadron duality in
individual helicity-1/2 and 3/2 virtual photoproduction cross sections.
We find that duality is realized more clearly in the helicity-1/2
channel, with duality violating corrections $\lesssim 10\%$ over
the entire nucleon resonance region, while larger, $\lesssim 20\%$
corrections are found in the helicity-3/2 sector.
The results are in general agreement with quark model expectations,
and suggest that data above the $\Delta$ resonance region may be used
to constrain both spin-averaged and spin-dependent parton distributions.
\end{abstract}

\maketitle

\section{Introduction}
\label{sec:intro}

The duality between quark and hadron descriptions of physical
observables reveals a fascinating connection between the physics
of quark confinement at low momentum scales and asymptotic freedom
at large momenta.
The striking manifestation of this duality in inclusive
electron--nucleon scattering, first observed \cite{BG} even before
the advent of QCD, has motivated considerable effort in recent years
to explore this phenomenon empirically, as well as to understand
its origins theoretically (for a review see Ref.~\cite{MEK}).

From a practical perspective, the quantitative demonstration of the
similarity between the structure functions measured in the nucleon
resonance region and those in the deep inelastic continuum, at higher
energies, opens up the intriguing possibility of using resonance
region data to provide constraints on leading twist parton
distribution functions (PDFs).
Attempts to utilize this connection have begun to be explored in
recent global PDF fits \cite{CTEQX} (see also Ref.~\cite{ABKM}),
where data on the unpolarized proton and deuteron $F_2$ structure
functions at final state hadronic masses $W$ as low as $\sim 1.7$~GeV
have been used to extend determinations of PDFs to larger values of
the Bjorken scaling variable $x = Q^2/2M\nu$, where $Q^2$ and $\nu$
are the four-momentum squared and energy transferred to the proton,
and $M$ is the proton mass.

The availability of high-luminosity electron beams at Jefferson Lab
has enabled high-precision measurements of various structure
functions to be made over the past decade.  These data have now
firmly established the existence of duality in the proton $F_2$
and $F_L$ structure functions \cite{IOANA,LIANG,TRUNC,MALACE},
and have provided tantalizing glimpses of its spin and flavor
dependence in polarized \cite{SPIN-p,SPIN-n} and semi-inclusive
scattering measurements \cite{SIDIS}.
Recently a new method \cite{KMK} was used to extract also the
neutron $F_2$ structure function from inclusive proton and
deuterium data in the nucleon resonance region \cite{MKMK},
leading to the first quantitative determination of duality
in the neutron's unpolarized structure functions.
This observation suggested that duality is indeed a general
feature of the resonance-scaling transition, and not due to
accidental cancellations of quark charges \cite{CHARGES}.

Because of the considerably larger data base of spin-averaged
cross sections than polarization asymmetries, duality in the
spin-dependent $g_1$ and $g_2$ structure functions has not yet
been established to the same precision as for the unpolarized
$F_{1,2}$ structure functions.
An additional complication arises from the fact that for
spin-dependent quantities one deals with differences of
cross sections, which are not restricted to be positive.
For example, in the $\Delta$ resonance region the $g_1$ structure
function of the proton, especially at low $Q^2$, is large and
negative, while for the same $x$ and higher $Q^2$
(hence higher $W^2 = M^2 + Q^2 (1-x)/x$) the structure function 
measured in the deep inelastic region is positive.
Such strong violation of duality would limit the use of resonance
region data to constrain spin-dependent PDFs.
Furthermore, the neutron $g_1$ structure function changes sign
as a function of $x$, so cannot be used to study ratios of
resonance to deep inelastic structure functions or the relative
size of duality violations.

On the other hand, duality violation may be less severe, even at low
$W$, in individual virtual photoabsorption helicity cross sections,
defined by projecting the total spin of the $\gamma^*$--proton center 
of mass system along the photon direction.
The helicity-1/2 projection, $\sigma_{1/2}$, represents the cross
section for equal initial and excited hadronic state helicities,
while the helicity-3/2 projection, $\sigma_{3/2}$, involves a change
of the hadron helicities by two units.
The sums and differences of the helicity cross sections, which are
positive definite, correspond to the unpolarized and spin polarized
structure functions, respectively.

At high $Q^2$ and $W^2$ the helicity cross sections are proportional
to the positive and negative helicity PDFs, $q^\pm(x,Q^2)$,
which describe the distribution of quarks with spin parallel or
antiparallel to that of the nucleon.
As with the cross sections, the helicity PDFs are defined to be
positive, and in a way represent more fundamental objects than
the spin-averaged and spin-dependent PDFs.
In fact, from perturbative QCD arguments one can make definite
predictions for the behavior of the helicity PDFs in the limit
$x \to 1$, with $q^-/q^+ \sim (1-x)^2$ \cite{FJ,LB}.
These predictions can be tested by studying the asymptotic $x$
dependence of helicity cross sections, which is difficult,
however, because of the rapidly decreasing rates as $x \to 1$.
Indeed, since large $x$ generally corresponds to low $W$,
determining the large-$x$ behavior of inclusive cross sections
at any finite $Q^2$ will necessarily involve the resonance region.
%

In this paper we perform the first detailed study of quark-hadron
duality in $\gamma^* p$ helicity cross sections, by combining
previously measured sets of data on inclusive spin-averaged cross
sections and double polarization asymmetries from Jefferson Lab
and elsewhere.
These data are used to quantify the degree of duality violation in
each of the three prominent nucleon resonance regions, as well as
over the entire range $W < 2$~GeV.

In Sec.~\ref{sec:defns} we begin by defining the relevant cross
sections and distributions used in this analysis.
Section~\ref{sec:data} outlines the data analysis, describing
the construction of the helicity cross sections form separate
measurements of spin-averaged and spin-dependent structure functions.
Results of the analysis are presented in Sec.~\ref{sec:results},
for the $x$ dependence of helicity structure functions in several
fixed-$Q^2$ bins ranging from $Q^2 = 1.7$ to 5~GeV$^2$, as well as
for integrals over the various resonance regions.
Comparison of the resonance data with parametrizations of data at
higher energies then allow the first determination of the extent to
which duality holds in helicity cross sections over this range.
We also compare our findings with quark models that predict
specific patterns of duality violation in structure functions.
Finally, in Sec.~\ref{sec:conc} we draw some conclusions from this
analysis and outline its broader implications for our understanding
of quark-hadron duality as well as its practical exploitation.

\section{Helicity structure functions}
\label{sec:defns}

The differential cross section for the inclusive scattering of a
longitudinally polarized electron with helicity $h=\pm 1$ from a
proton with polarization $P_z$ along the virtual photon direction
can be written as
\begin{eqnarray}
{ d\sigma \over d\Omega dE' }
&=& \Gamma
    \left( \sigma_T + \varepsilon \sigma_L
	 + h P_z \sqrt{1-\varepsilon^2} \sigma'_{TT}
    \right),
\end{eqnarray}
where $E'$ is the scattered electron energy, $\Gamma$ is the flux
of virtual photons and $\varepsilon$ is the transverse photon
polarization \cite{DRECHSEL}.
The photoabsorption cross sections for transversely polarized
virtual photons are related to the helicity cross sections
$\sigma_{1/2}$ and $\sigma_{3/2}$ by
\begin{eqnarray}
\sigma_T
&=& {1\over 2} \left( \sigma_{1/2} + \sigma_{3/2} \right),	\\
\sigma'_{TT}
&=& {1\over 2} \left( \sigma_{3/2} - \sigma_{1/2} \right),
\end{eqnarray}
while the cross section for longitudinally polarized photons
is given by the longitudinal structure function.
The cross section $\sigma_{1/2}$ ($\sigma_{3/2}$) corresponds to the
spins of the virtual photon and proton antialigned (aligned) in the
center of mass system, so that the helicity of the excited nucleon
state after absorbing a photon is $+1/2$ ($+3/2$).  Whereas the
$\sigma_{1/2}$ cross section conserves the nucleon helicity, the
$\sigma_{3/2}$ changes the nucleon helicity by two units.

For convenience we define dimensionless helicity structure functions
$H_{1/2}$ and $H_{3/2}$ in terms of the cross sections by
\begin{equation}
H_{1/2}\ =\ {M K \over 4\pi^2\alpha}\, \sigma_{1/2},\ \ \ \
H_{3/2}\ =\ {M K \over 4\pi^2\alpha}\, \sigma_{3/2},
\end{equation}
where $\alpha = e^2/4\pi$ and $K = (W^2-M^2)/2M$ is associated with the
choice of the virtual photon flux in the Hand convention \cite{HAND}.
The helicity structure functions can then be written in terms of the
usual unpolarized $F_1$ and polarized $g_{1,2}$ structure functions as
\begin{subequations}
\label{eq:H}
\begin{eqnarray}
H_{1/2} &=& F_1 + g_1 - {Q^2 \over \nu^2} g_2, 	\\
%
H_{3/2} &=& F_1 - g_1 + {Q^2 \over \nu^2} g_2,
\end{eqnarray}
\end{subequations}
each of which is a function of two variables, typically taken to be
$x$ and $Q^2$.
In the limit where both $Q^2$ and $W^2$ are large, with $x$ finite,
the $F_1$ and $g_1$ structure functions can be written, at leading
order in $\alpha_s$, in terms of leading twist PDFs,
\begin{subequations}
\begin{eqnarray}
F_1 &=& {1 \over 2} \sum_q e_q^2\, (q + \bar q),		\\
g_1 &=& {1 \over 2} \sum_q e_q^2\, (\Delta q + \Delta\bar q),
\end{eqnarray}
\end{subequations}
where $q = q^+ + q^-$ and $\Delta q = q^+ - q^-$ are the spin-averaged
and spin-dependent PDFs.  In this case the helicity structure functions
become
\begin{subequations}
\begin{eqnarray}
H_{1/2} &=& \sum_q e_q^2\, (q^+ + \bar q^+),		\\
H_{3/2} &=& \sum_q e_q^2\, (q^- + \bar q^-),
\end{eqnarray}
\end{subequations}
so that in this limit $H_{1/2}$ is determined by the $q^+$ PDFs
while $H_{3/2}$ is determined by the $q^-$ PDFs (the antiquark
distributions $\bar q^+$ and $\bar q^-$ are suppressed by
additional powers of $(1-x)$ compared with the quark PDFs).
In the $x \to 1$ limit the leading behavior of the helicity
distributions is predicted from perturbative QCD to be
$q^+ \sim (1-x)^3$ and $q^- \sim (1-x)^5$ if the nucleon
ground state wave function is dominated by its $S$-wave
component \cite{FJ,LB}, or $q^- \sim (1-x)^5 \log^2(1-x)$
if one includes orbital angular momentum \cite{ABDY}.

\section{Data analysis}
\label{sec:data}

The experimental $H_{1/2}$ and $H_{3/2}$ helicity structure functions
used in this analysis were obtained by combining measurements of
$g_1/F_1$ ratios from Jefferson Lab experiment EG1b (E91-023) in CLAS
\cite{EG1b} with the unpolarized $F_1$ structure function from the
empirical Christy-Bosted (CB) global fit \cite{CHRISTY}.
For the small correction from the $g_2$ structure function we use
the phenomenological parametrization of Ref.~\cite{SIMULA}.

The CLAS E91-023 data set represents one of the few high-precision
measurements of $g_1/F_1$ for the proton at moderate to large $x$
($x > 0.15$), covering a large $Q^2$ range from 0.05 to 5 GeV$^2$,
over both the resonance and DIS regions.
The empirical CB global fit uses measurements of inclusive
inelastic electron-proton cross sections in the kinematic range
of $Q^2 < 8$~GeV$^2$ and $W$ between 1.1 and 3.1~GeV. 
The fit is constrained by high precision longitudinal and transverse
separated cross section measurements from Jefferson Lab Hall~C
\cite{LIANG}, unseparated Hall~C measurements up to $Q^2$ of
7.5~GeV$^2$ \cite{MALACE}, and photoproduction data at $Q^2 = 0$.
This fit was chosen because it covers a wide kinematic range and
uses both transverse and longitudinal cross sections, which is
particularly important for the $F_1$ estimation.
Due to the scarcity of $g_2$ measurements, especially in the
resonance region, we use the phenomenological parametrization of
Ref.~\cite{SIMULA}, which is developed for $x > 0.02$ using DIS
data with $Q^2$ up to 50~GeV$^2$ as well as experimental results
on both photo- and electroproduction of proton resonances.

The statistical uncertainties for $H_{1/2}$ and $H_{3/2}$ were
calculated from those of the $g_1/F_1$ measurements \cite{EG1b}.
The systematic uncertainties were obtained from those of $g_1/F_1$,
$F_1$ and $g_2$ by varying these quantities within the limits given
by their systematics.
The resulting variations of $H_{1/2}$ and $H_{3/2}$ were then
added in quadrature to obtain the total systematic uncertainty.
Note that since $H_{1/2}$ and $H_{3/2}$ are different combinations
of the same structure functions (differing only in relative sign,
Eqs.~(\ref{eq:H})), the resulting absolute uncertainties are
the same for the two cross sections.
However, since $H_{3/2} \ll H_{1/2}$ the relative uncertainties will
be different, with that on $H_{3/2}$ much greater than on $H_{1/2}$.

The theoretical $H_{1/2}$ and $H_{3/2}$ structure functions were
obtained by combining $g_1$ from the Bl\"umlein-B\"ottcher (BB) global
parametrization \cite{BB} of spin-dependent structure functions with
$F_1$ constructed from the $F_2$ global fit of Alekhin {\it et al.}
\cite{ABKM} and the R1998 parametrization of the longitudinal to
transverse cross section ratio $R$ \cite{R1998}.
The BB global fit \cite{BB} is based on a next-to-leading order QCD
analysis of the world data on polarized deep inelastic scattering,
and includes possible higher twist contributions.
The analysis finds that for both proton and deuteron targets the
higher twist corrections to $g_1$ are consistent with zero,
within the large uncertainties of the data.
For the $g_2$ structure function we therefore use the Wandzura-Wilczek
relation \cite{WW} with $g_1$ from the BB fit.

The global fit of Alekhin {\it et al.} \cite{ABKM} provides QCD
parametrizations for both the $F_1$ and $F_2$ structure functions.
While the $F_2$ fit reproduces well the available $F_2$ data,
the $F_1$ fit shows some discrepancies at low $W$ with the
high-precision longitudinal and transverse separated cross section
measurements from Jefferson Lab Hall~C \cite{LIANG}.
We find that a good description of the Hall~C $F_1$ data can be
obtained by using the $F_2$ fit from Ref.~\cite{ABKM} with the
R1998 parametrization of $R$.

\section{Results}
\label{sec:results}

\begin{figure}[ht]
\includegraphics[width=14cm]{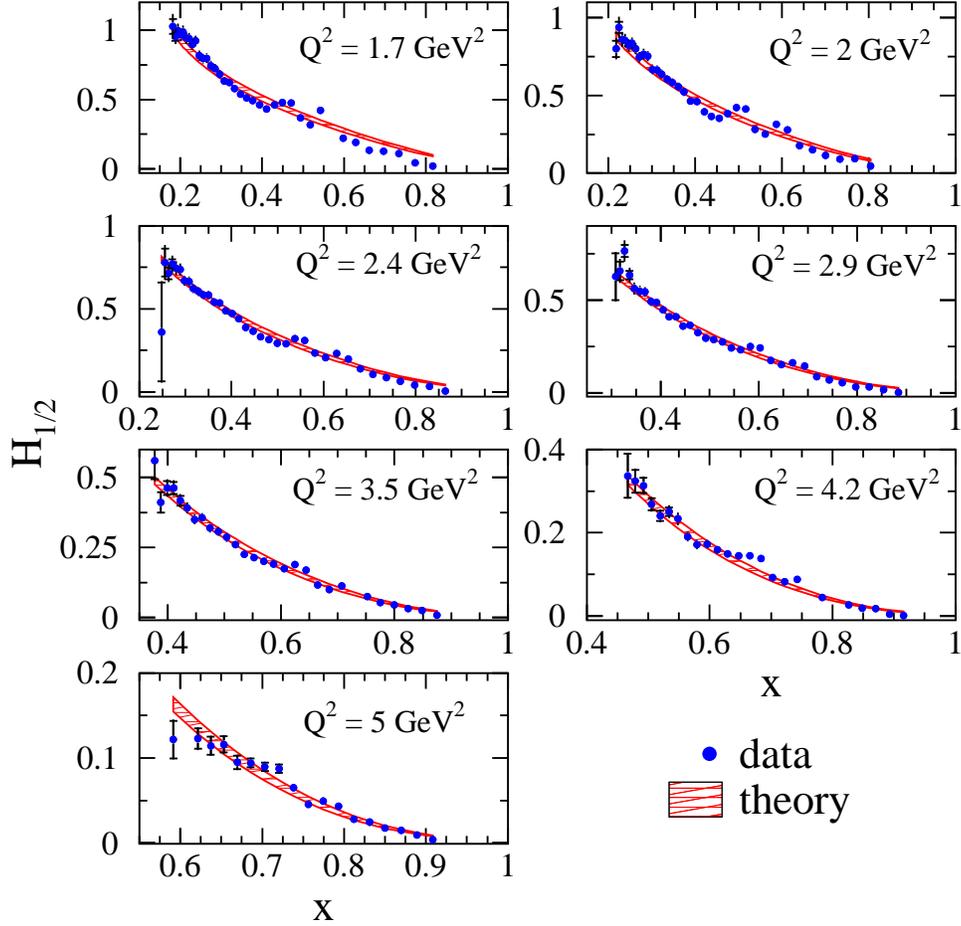}
\caption{
	Scaled helicity-$1/2$ cross section $H_{1/2}$
	as a function of $x$ for various $Q^2$ bins.
	The bands (labeled ``theory'') represent a global fit to
	high-$W$ data (see text).}
\label{fig:H12}
\end{figure}

\begin{figure}[ht]
\includegraphics[width=14cm]{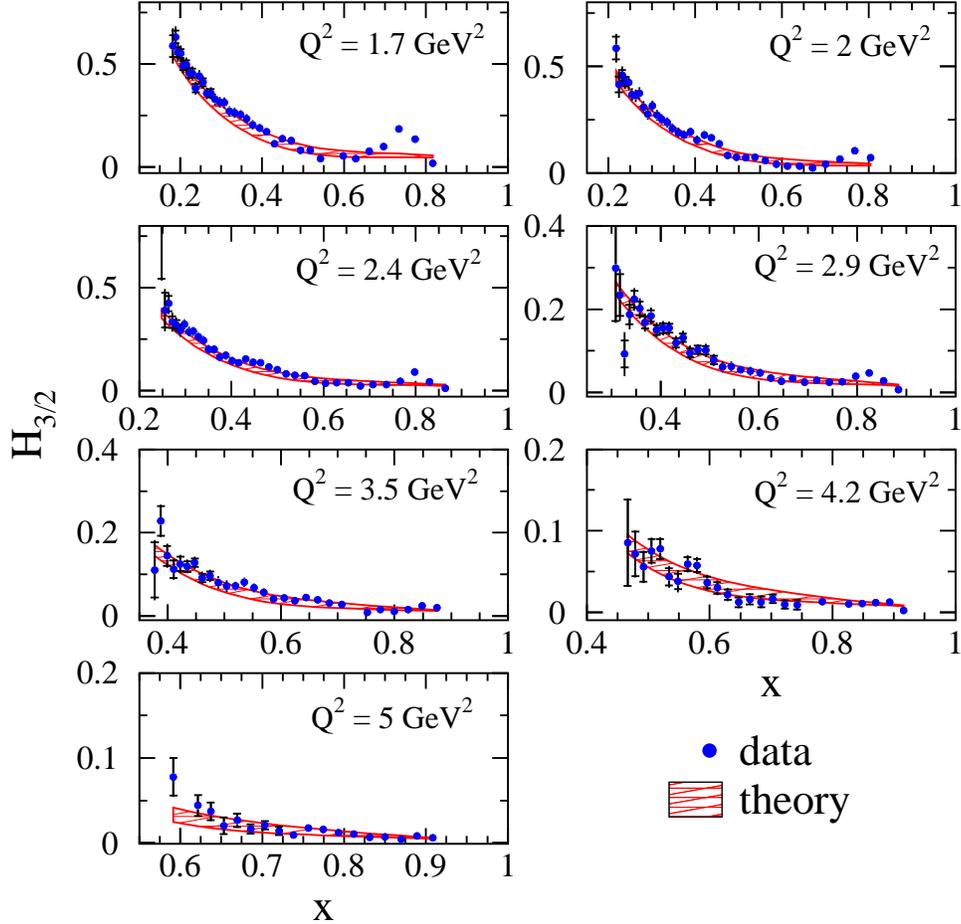}
\caption{
	As in Fig.~\ref{fig:H12} but for the helicity-$3/2$
	cross section $H_{3/2}$.}
\label{fig:H32}
\end{figure}

The results for the helicity structure functions $H_{1/2}$ and $H_{3/2}$
are shown in Figs.~\ref{fig:H12} and \ref{fig:H32} as a function of $x$
for several fixed $Q^2$ values ranging from $Q^2=1.7$~GeV$^2$ to
5~GeV$^2$.  The data are compared with curves (labeled ``theory'')
constructed from global fits to structure functions in the deep inelastic
region at higher $W$, as outlined in Sec.~\ref{sec:data} above.
The resonance region data are in excellent agreement with the global fit
for the $H_{1/2}$ structure function for the kinematics considered.
The agreement for $H_{3/2}$ is also quite good overall, although here
the ``theory'' curve slightly underestimates the data, especially at
lower $Q^2$ and in the $\Delta$ resonance region, where a prominent peak
stands out.  This can be understood from the fact that at low $Q^2$ the
$\Delta$ contribution to $F_1$ is positive, while that to $g_1$ is
negative, thereby cancelling in $H_{1/2}$ but reinforcing in $H_{3/2}$.


\begin{figure}
\includegraphics[width=14cm]{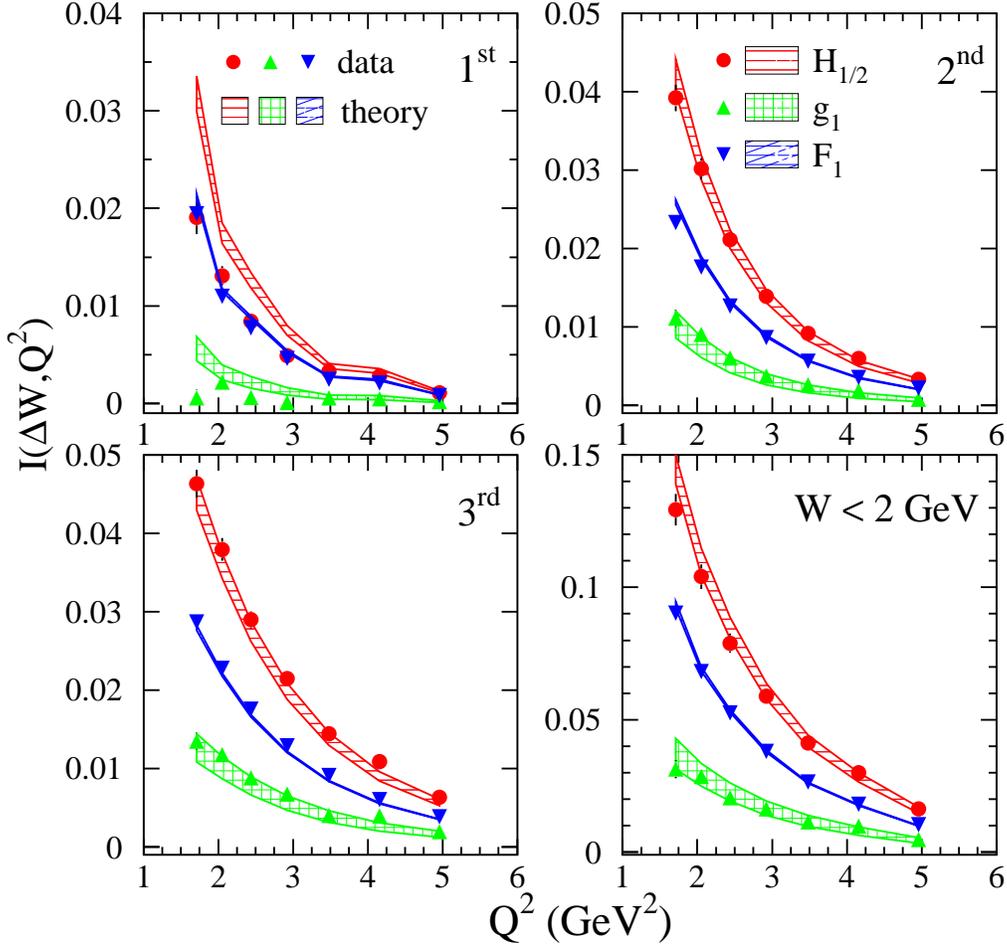}
\caption{
	Integrals $I(\Delta W,Q^2)$ of the scaled helicity-$1/2$
	structure function $H_{1/2}$ in various resonance regions
	$\Delta W$ (1st, 2nd, 3rd and $W<2$~GeV) versus $Q^2$.
	For comparison the corresponding integrals of the $g_1$ and
	$F_1$ structure functions are also shown.
	The bands represent a global fit to high-$W$ data (see text).}
\label{fig:I12}
\end{figure}
\begin{figure}
\includegraphics[width=14cm]{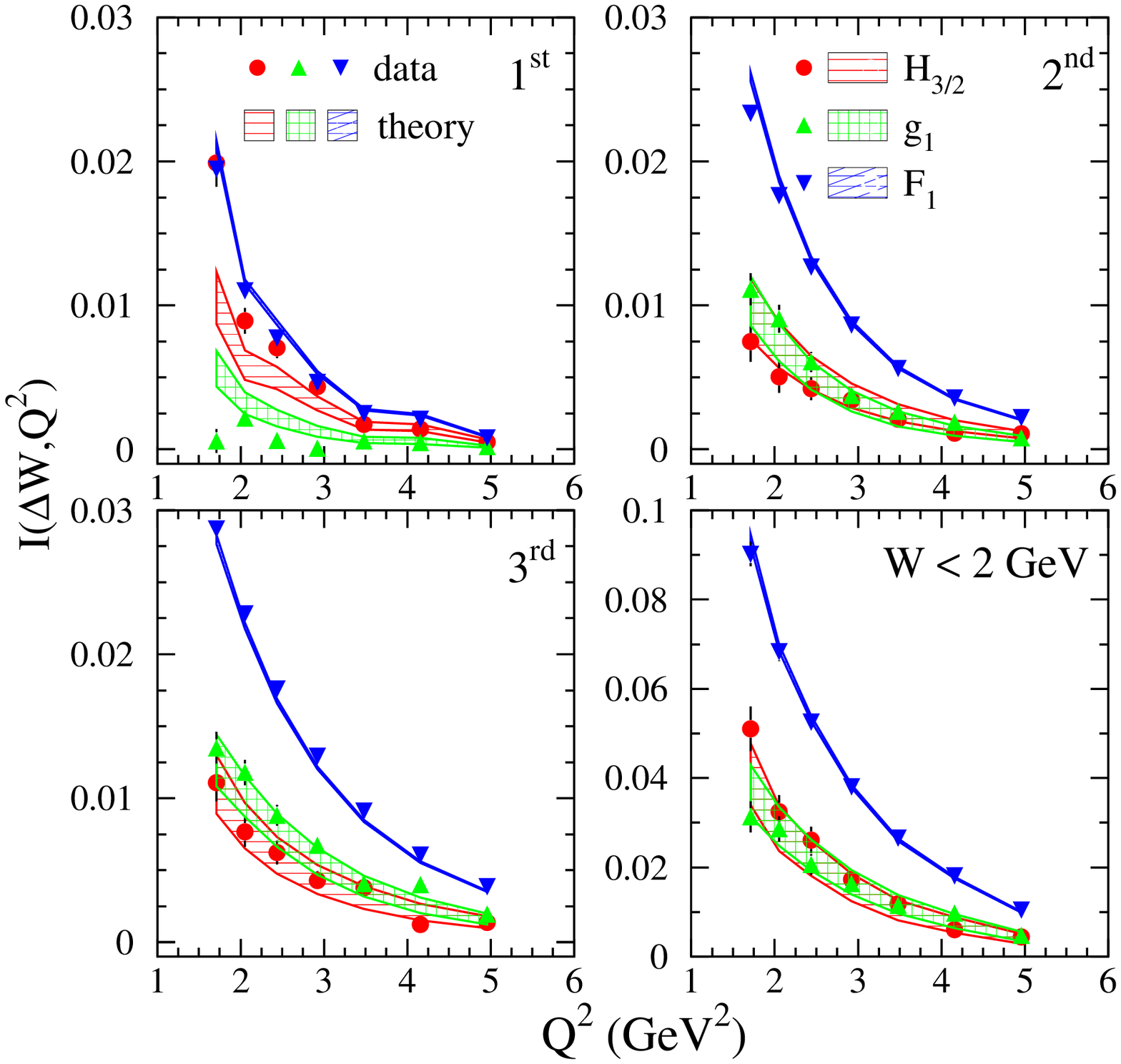}
\caption{
	As in Fig.~\ref{fig:I12} but for the helicity-$3/2$
	structure function $H_{3/2}$.}
\label{fig:I32}
\end{figure}

The degree to which duality holds can be quantified by considering
integrals of the structure functions over individual resonance regions,
$\Delta W$,
\begin{eqnarray}
I(\Delta W,Q^2)
&=& \int_{\Delta W} dx\, {\cal F}(x,Q^2),
\end{eqnarray}  
where ${\cal F} = H_{1/2}, H_{3/2}, F_1$ or $g_1$.
Following earlier data analyses \cite{IOANA,MALACE,MKMK}, we take for
$\Delta$ the three prominent resonance regions, defined on the intervals
\begin{itemize}

\item	$1^{\rm st}$ resonance region: $1.3 \leq W^2 \leq 1.9$~GeV$^2$

\item	$2^{\rm nd}$ resonance region: $1.9 \leq W^2 \leq 2.5$~GeV$^2$

\item	$3^{\rm rd}$ resonance region: $2.5 \leq W^2 \leq 3.1$~GeV$^2$

\end{itemize}
as well as the entire resonance region $W^2 \leq 4$~GeV$^2$.
These are shown in Figs.~\ref{fig:I12} and \ref{fig:I32} for the
$H_{1/2}$ and $H_{3/2}$ cases, respectively, with the integrals
for $F_1$ and $g_1$ shown in comparison.
Because $H_{1/2}$ involves a sum of the (positive) $F_1$ and
(generally positive) $g_1$ structure functions, the $H_{1/2}$
data are generally larger in magnitude than $F_1$ and $g_1$.
Above the first resonance region the agreement with the global
fits (shaded regions) is extremely good over the entire range
of $Q^2$ considered for each of the $H_{1/2}$, $F_1$ and $g_1$
structure functions.
The agreement in the first resonance region is markedly
worse, reflecting the strong violation of duality in the $g_1$
structure function in the vicinity of the $\Delta$ resonance.
This violation persists until $Q^2 \approx 3$~GeV$^2$, above which
the resonance and deep inelastic data are in better agreement.

\begin{figure}
\includegraphics[width=14cm]{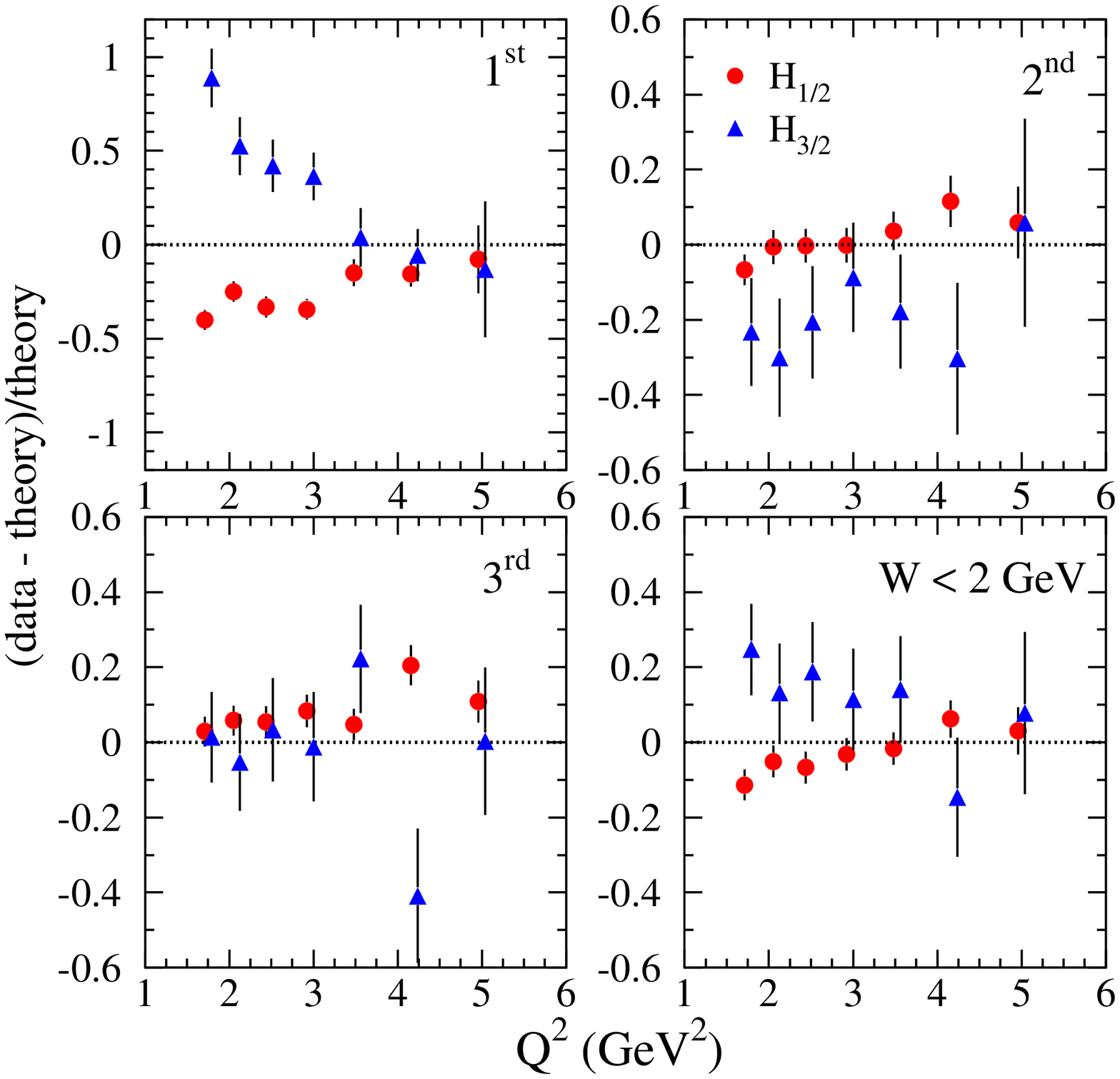}
\caption{
	Ratio (data -- theory)/theory for the scaled helicity cross
	sections $H_{1/2}$ and $H_{3/2}$ in various resonance regions
	$\Delta W$ (1st, 2nd, 3rd and $W<2$~GeV) versus $Q^2$.
	Some of the points have been offset for clarity.}
\label{fig:rel}
\end{figure}

For the $H_{3/2}$ structure function, because this involves the
difference between $F_1$ and $g_1$, its magnitude is considerably
smaller than that of $F_1$.  Again, duality violation is strongest
in the $\Delta$ region, with generally good agreement between
resonance and deep inelastic data at higher $W$.
While the violations of duality are expected to diminish at larger
$Q^2$, the decreasing magnitude of the higher-$Q^2$ integrals makes
it more difficult to quantify the violation accurately.

To ameliorate this problem we compute the ratios of the integrals
of the resonance region data to those of the global fits, shown in
Fig.~\ref{fig:rel} for $H_{1/2}$ and $H_{3/2}$.
In general these ratios show that duality violation is stronger in
the helicity-3/2 channel than in the helicity-1/2, with the duality
violating corrections for $H_{3/2}$ positive in the first resonance 
region and negative in the second resonance region.
As could be expected, the differences between the data and theory
are largest in the $\Delta$ region for both $H_{1/2}$ and $H_{3/2}$.
The larger uncertainties on the $H_{3/2}$ data reflects the fact
that $H_{3/2} \ll H_{1/2}$.
Integrating over the entire $W < 2$~GeV region, the duality violation
is $\lesssim 10\%$ for $H_{1/2}$ and $\lesssim 20\%$ for $H_{3/2}$
at $Q^2 \leq 4$~GeV$^2$.
This is considerably smaller than the corresponding duality violation
found in the spin-dependent $g_1$ structure function \cite{SPIN-p}.

\begin{table}[h]
\caption{Relative strengths of $N \to N^*$ transitions for helicity
	structure functions $H_{1/2, 3/2}$ in the SU(6) quark model
	\cite{CI}.
	The coefficients $\lambda$ and $\rho$ denote the relative
	strengths of the symmetric and antisymmetric contributions
	of the SU(6) ground state wave function, with the SU(6) limit
	corresponding to $\lambda = \rho$.\\}
\begin{tabular}{c|ccccc|c}            \hline
SU(6) rep.      & $^2${\bf 8}[{\bf 56}$^+$]\ \
                & $^4${\bf 10}[{\bf 56}$^+$]\ \
                & $^2${\bf 8}[{\bf 70}$^-$]\ \
                & $^4${\bf 8}[{\bf 70}$^-$]\ \
                & $^2${\bf 10}[{\bf 70}$^-$]\ \
                & total\ \                              \\ \hline
$H_{1/2}$ & $9 \rho^2$
          & $2 \lambda^2$
          & $9 \rho^2$
          & $0$
          & $\lambda^2$   
          & $18 \rho^2 + 3 \lambda^2$                     \\
$H_{3/2}$ & 0
          & $6 \lambda^2$
          & 0
          & 0
          & 0
          & $6 \lambda^2$		\\ \hline
\end{tabular}
\vspace*{0.5cm}
\end{table}

Our results can be compared with quark model predictions for the
relative strengths of the $N \to N^*$ transitions.
In Tab.~1 these are displayed for $H_{1/2}$ and $H_{3/2}$ in the
various SU(6)$^P$ = {\bf 56}$^+$ ($L=0$) and {\bf 70}$^-$ ($L=1$)
representations \cite{CI,CM}, with each representation weighted
equally.
The contributions from the symmetric and antisymmetric components
of the ground state nucleon wave function enter with strengths
$\lambda$ and $\rho$, respectively, and the SU(6) limit corresponds
to $\lambda = \rho$.
The usual quark model assignments of the excited states have the
nucleon and $\Delta$ in the quark spin-${1\over 2}$ $^2${\bf 8} and
quark spin-${3 \over 2}$ $^4${\bf 10} representations of {\bf 56}$^+$,
respectively.
For the odd parity states the $^2${\bf 8} multiplet contains the
states $S_{11}(1535)$ and $D_{13}(1520)$, which dominate the second
resonance region; the $^4${\bf 8} contains the $S_{11}(1650)$,
$D_{13}(1700)$ and $D_{15}(1675)$; and the isospin-${3 \over 2}$
states $S_{31}(1620)$ and $D_{33}(1700)$ belong to the $^2${\bf 10}
representation \cite{SU6}.

With the exception of the $\Delta$ region, the $H_{1/2}$ structure
function is predicted to be much larger than the $H_{3/2}$, as is
borne out by the data in Figs.~\ref{fig:H12} and \ref{fig:H32}.
The relatively small contribution to $H_{1/2}$ in the
$^4${\bf 10}[{\bf 56}$^+$] channel and large contribution in the
$^2${\bf 8}[{\bf 70}$^-$] channel suggests that the helicity-1/2
data should lie below the global fit in the first resonance region
and above the global fit at larger $W$.
This is generally consistent with the data in Fig.~\ref{fig:rel}.

The helicity-3/2 structure function is dominated in the resonance region
by the $\Delta$, with suppressed contributions in all other channels.
Again this is consistent with the $H_{3/2}$ data being higher than
the global fit in the $\Delta$ region and below the fit at larger $W$.
The prediction of vanishing $H_{3/2}$ for the nucleon elastic
contribution reflects the dominance of magnetic coupling assumed in
the model \cite{SU6}, which is expected to be a better approximation
at high $Q^2$.

Summing over all channels, the ratio of helicity-3/2 to 1/2 structure
functions is predicted to be $H_{3/2}/H_{1/2} = 2/7$, which coincides
exactly with the quark-parton model results
$u^-/u^+ = 1/5 = d^+/u^+$ and $d^-/d^+ = 2 = d^-/u^-$ for all $x$.
These predictions are found to hold approximately at $x \sim 1/3$,
but significant deviations are observed at larger $x$.
Various scenarios for SU(6) symmetry breaking, consistent with
quark-hadron duality, were considered in Ref.~\cite{CM}, leading to
specific predictions for structure function ratios in the $x \to 1$
limit.
The general trends of the duality violations persist even in the more
realistic symmetry breaking scenarios, so that the deviations from unity
in Fig.~\ref{fig:rel} can be understood, at least qualitatively, in
terms of a microscopic quark-level description.
Note also that the quark model predictions relate to the resonant
components of the data {\it only}; the presence of the nonresonant
background washes out these predictions somewhat, especially at larger
$Q^2$, and its remarkable that the general trends of the duality
violations in the various resonance regions nevertheless remain.

\section{Conclusions}
\label{sec:conc}

In this work we have performed the first detailed analysis of
quark-hadron duality in individual $\gamma^* p$ helicity cross
sections, utilizing recent data on inclusive unpolarized and
polarized structure functions from Jefferson Lab.
Unlike spin-dependent structure functions which can change sign
as a function of $x$ and $Q^2$, the helicity cross sections are
by definition constrained to be positive definite.
This reduces the dramatic violations of duality seen for example
in the proton $g_1$ structure function in the $\Delta$ resonance
region, where the negative resonance contribution at low $Q^2$
makes way for a positive structure function in deep inelastic
kinematics at large $Q^2$.

The data on the polarized and unpolarized structure functions are
used to quantify the degree of duality violation in the helicity
structure functions in each of the three prominent nucleon resonance
regions, as well as over the entire range $W < 2$~GeV.
We find that duality is realized more clearly for helicity-1/2
structure function $H_{1/2}$ than for the helicity-3/2 function
$H_{3/2}$, with the duality violating corrections in the latter
positive in the first resonance region and negative in the second
resonance region.  The duality violations are largest in the $\Delta$
region for both $H_{1/2}$ and $H_{3/2}$.

Over entire resonance region, the duality violating corrections are
$\lesssim 10\%$ (and negative) for $H_{1/2}$ and $\lesssim 20\%$
(and positive) for $H_{3/2}$ at $Q^2 \leq 4$~GeV$^2$,
which is rather smaller than the corresponding duality violation
found in the spin-dependent $g_1$ structure function.
The patterns of duality violation are in general agreement with
expectations from quark models based on spin-flavor symmetry
\cite{CI,CM}.

Our results suggest that data above the $\Delta$ resonance region
could be used to constrain {\it both} spin-averaged and spin-dependent
parton distributions.  This lends support to recent efforts to
broaden the kinematic coverage in global fits of unpolarized PDFs
\cite{CTEQX} by lowering the $Q^2$ and $W^2$ cuts, and to extending
these efforts to the polarized sector.
Moreover, it raises the interesting possibility of performing global
fits of helicity PDFs $q^+$ and $q^-$ directly, rather than
reconstructing these from separate unpolarized and polarized PDF
analyses; such an enterprise would demand a consistent analysis of
combined cross section and polarization asymmetry data along the
lines presented in this work.

\begin{acknowledgments}

This work was supported by the U.S. Department of Energy under Contract
No. DE-FG02-03ER41231, and DOE contract No. DE-AC05-06OR23177, under
which Jefferson Science Associates, LLC operates Jefferson Lab.

\end{acknowledgments}


\end{document}